\documentclass[useAMS,usenatbib]{mn2e}
\usepackage{mn2e-breakabs}
\usepackage{graphicx}
\usepackage{subfigure}
\usepackage{times}
\usepackage{caption}
\usepackage{rotating}
\usepackage{rotfloat}
\usepackage{multicol}
\usepackage{array}
\usepackage{booktabs}
\usepackage{color}
\usepackage{afterpage}
\newcommand{\Hunit}{\,{\rm km}\,{\rm s}^{-1}\,{\rm Mpc}^{-1}}

\def\la{\mathrel{\mathpalette\fun <}}
\def\ga{\mathrel{\mathpalette\fun >}}
\def\fun#1#2{\lower3.6pt\vbox{\baselineskip0pt\lineskip.9pt
        \ialign{$\mathsurround=0pt#1\hfill##\hfil$\crcr#2\crcr\sim\crcr}}}

\newcommand{\be}{\begin{equation}}
\newcommand{\ee}{\end{equation}}
\newcommand{\ba}{\begin{eqnarray}}
\newcommand{\ea}{\end{eqnarray}}
\newcommand{\simgt}{\,\hbox{\lower0.6ex\hbox{$\sim$}\llap{\raise0.6ex\hbox{$>$}}}\,}
\newcommand{\simlt}{\,\hbox{\lower0.6ex\hbox{$\sim$}\llap{\raise0.6ex\hbox{$<$}}}\,}

\begin{document}

\title[Single-Probe Measurements from SDSS BOSS CMASS]
{The clustering of galaxies in the SDSS-III Baryon Oscillation Spectroscopic Survey:
single-probe measurements from CMASS anisotropic galaxy clustering
}

\author[Chuang et al.]{
  \parbox{\textwidth}{
 Chia-Hsun Chuang$^{1,2}$\thanks{MultiDark Fellow; achuang@aip.de},
 Francisco Prada$^{1,3,4}$, 
 Marcos Pellejero-Ibanez$^{5,6}$,
Florian Beutler$^{7,8}$,
Antonio J. Cuesta$^{9}$,
Daniel J. Eisenstein$^{10}$,
Stephanie Escoffier$^{11}$,
Shirley Ho$^{12}$,
Francisco-Shu Kitaura$^{2,7,13}$,
Jean-Paul Kneib$^{14,15}$,
Marc Manera$^{16,8}$,
Sebasti\'an E. Nuza$^{2}$,
Sergio~Rodr\'iguez-Torres$^{1,3,17}$,
Ashley Ross$^{18}$,
J. A. Rubi\~no-Mart\'{\i}n$^{5,6}$,
Lado Samushia$^{19,20,8}$,
David J. Schlegel$^{7}$,
Donald P. Schneider$^{21,22}$,
Yuting Wang$^{23,8}$,
Benjamin A. Weaver$^{24}$,
Gongbo Zhao$^{23,8}$,
Joel R. Brownstein$^{25}$,
Kyle S. Dawson$^{25}$,
Claudia Maraston$^{8}$,
Matthew D Olmstead$^{26}$,
Daniel Thomas$^{8}$
}
  \vspace*{4pt} \\
$^1$ Instituto de F\'{\i}sica Te\'orica, (UAM/CSIC), Universidad Aut\'onoma de Madrid,  Cantoblanco, E-28049 Madrid, Spain \\
$^{2}$ Leibniz-Institut f\"ur Astrophysik Potsdam (AIP), An der Sternwarte 16, 14482 Potsdam, Germany\\
$^3$ Campus of International Excellence UAM+CSIC, Cantoblanco, E-28049 Madrid, Spain \\ 
$^4$ Instituto de Astrof\'{\i}sica de Andaluc\'{\i}a (CSIC), Glorieta de la Astronom\'{\i}a, E-18080 Granada, Spain \\
$^{5}$ Instituto de Astrof\'isica de Canarias (IAC), C/V\'ia L\'actea, s/n, E-38200, La Laguna, Tenerife, Spain\\
$^{6}$ Departamento Astrof\'isica, Universidad de La Laguna (ULL), E-38206 La Laguna, Tenerife, Spain\\
$^7$ Lawrence Berkeley National Lab, 1 Cyclotron Rd, Berkeley CA 94720, USA\\ 
$^{8}$ Institute of Cosmology and Gravitation, University of Portsmouth, Dennis Sciama Building, Portsmouth PO1 3FX, UK\\ 
$^{9}$ Institut de Ci{\`e}ncies del Cosmos (ICCUB), Universitat de Barcelona (IEEC-UB), Mart{\'\i} i Franqu{\`e}s 1, E08028 Barcelona, Spain\\
$^{10}$ Harvard-Smithsonian Center for Astrophysics, 60 Garden St., Cambridge, MA 02138, USA\\ 
$^{11}$ CPPM, Aix-Marseille Universit\'{e}, CNRS/IN2P3, Marseille, France\\ 
$^{12}$ Department of Physics, Carnegie Mellon University, 5000 Forbes Ave., Pittsburgh, PA 15213, USA\\ 
$^{13}$ Departments of Physics and Astronomy, University of California, Berkeley, CA 94720, USA\\ 
$^{14}$ Laboratoire d'astrophysique, \'Ecole Polytechnique F\'ed\'erale de Lausanne (EPFL), Observatoire de Sauverny, 1290 Versoix, Switzerland\\ 
$^{15}$ Aix Marseille Universit\'e, CNRS, LAM (Laboratoire d'Astrophysique de Marseille) UMR 7326, 13388, Marseille, France\\ 
$^{16}$ Department of Physics \& Astronomy, University College London, Gower Street, London, WC1E 6BT,  UK\\ 
$^{17}$ Departamento de F\'isica Te\'orica M8, Universidad Autonoma de Madrid (UAM), Cantoblanco, E-28049, Madrid, Spain \\
$^{18}$ Center for Cosmology and Astroparticle Physics, Department of Physics, The Ohio State University, OH 43210, USA\\ 
$^{19}$ Kansas State University, Manhattan KS 66506, USA\\  
$^{20}$National Abastumani Astrophysical Observatory, Ilia State University, 2A Kazbegi Ave., GE-1060 Tbilisi, Georgia\\ 
$^{21}$ Department of Astronomy and Astrophysics, The Pennsylvania State University, University Park, PA 16802, USA\\ 
$^{22}$ Institute for Gravitation and the Cosmos, The Pennsylvania State University, University Park, PA 16802, USA\\ 
$^{23}$ National Astronomy Observatories, Chinese Academy of Science, Beijing, 100012, P.R.China\\
$^{24}$ Center for Cosmology and Particle Physics, New York University, New York, NY 10003 USA\\ 
$^{25}$ Department of Physics and Astronomy, University of Utah, 115 S 1400 E, Salt Lake City, UT 84112, USA\\
$^{26}$ Department of Chemistry and Physics, King's College, 133 North River St, Wilkes Barre, PA 18711, USA\\ 
}

\date{\today} 

\maketitle

\begin{abstract}
With the largest spectroscopic galaxy survey volume drawn from the SDSS-III Baryon Oscillation Spectroscopic Survey (BOSS),
we can extract cosmological constraints from the measurements of redshift and geometric distortions 
at quasi-linear scales (e.g. above 50 $h^{-1}$Mpc). 
We analyze the broad-range shape of the monopole
and quadrupole correlation functions of the BOSS Data Release 12 (DR12) CMASS galaxy sample, at the effective redshift $z=0.59$,
to obtain constraints on the Hubble expansion rate $H(z)$, the angular-diameter distance $D_A(z)$, 
the normalized growth rate $f(z)\sigma_8(z)$, and the physical matter density $\Omega_mh^2$. 
We obtain robust measurements by including a polynomial as the model for the systematic errors, and find it works very well 
against the systematic effects, e.g., ones induced by stars and seeing. 
We provide accurate measurements 
$\{D_A(0.59)r_{s,fid}/r_s$, $H(0.59)r_s/r_{s,fid}$, $f(0.59)\sigma_8(0.59)$, $\Omega_m h^2\}$ = $\{1427\pm26$ $\rm Mpc$, $97.3\pm3.3$ $\Hunit$, $0.488 \pm 0.060$, $0.135\pm0.016\}$, 
where $r_s$ is the comoving sound horizon at the drag epoch and $r_{s,fid}=148.66$ Mpc is the sound scale of the fiducial cosmology used in this 
study. 
The parameters which are not well constrained by our galaxy clustering analysis are marginalized over with wide flat priors. 
Since no priors from other data sets, e.g., cosmic microwave background (CMB), are adopted and no dark energy models are assumed, 
our results from BOSS CMASS galaxy clustering alone may be combined with other data sets, 
i.e., CMB, SNe, lensing or other galaxy clustering data to constrain the parameters of a given cosmological model.  
The uncertainty on the dark energy equation of state parameter, $w$, from CMB+CMASS is about 8 per cent. The uncertainty on the curvature fraction, $\Omega_k$, is 0.3 per cent. We do not find deviation from flat $\Lambda$CDM.
\end{abstract}

\begin{keywords}
 cosmology: observations - distance scale - large-scale structure of
  Universe - cosmological parameters
\end{keywords}

\section{Introduction} \label{sec:intro}

The cosmic large-scale structure from galaxy redshift surveys provides a 
powerful probe of dark energy and the cosmological model
that is highly complementary to the cosmic microwave 
background (CMB) (e.g., \citealt{Hinshaw:2012fq,Ade:2013ktc}), supernovae (SNe) 
\citep{Riess:1998cb,Perlmutter:1998np}, and weak lensing (e.g., see \citealt{VanWaerbeke:2003uq} for a review).

The scope of galaxy redshift 
surveys has dramatically increased in the last decade. The 2dF Galaxy Redshift Survey (2dFGRS) 
obtained 221,414 galaxy redshifts at $z<0.3$ \citep{Colless:2001gk,Colless:2003wz}, 
and the Sloan Digital Sky Survey (SDSS, \citealt{York:2000gk}) collected 
930,000 galaxy spectra in the Seventh Data Release (DR7) at $z<0.5$ \citep{Abazajian:2008wr}.
WiggleZ collected spectra of 240,000 emission-line galaxies at $0.5<z<1$ over 
1000 square degrees \citep{Drinkwater:2009sd, Parkinson:2012vd}, and the Baryon Oscillation Spectroscopic Survey (BOSS, \citealt{Dawson13}) of the SDSS-III \citep{Eisenstein11} is surveying 1.5 million luminous red galaxies (LRGs) at $0.1<z<0.7$ over 10,000 square degrees.
The newest BOSS data set has been made publicly available in SDSS data release 12 (DR12, \citealt{Alam:2015mbd}).
The planned space mission Euclid\footnote{http://sci.esa.int/euclid} will survey over 60 million emission-line galaxies at $0.7<z<2$ over 15,000 deg$^2$ 
(e.g. \citealt{RB}), and the upcoming ground-based experiment DESI\footnote{http://desi.lbl.gov/} (Dark Energy Spectroscopic Instrument) will survey 20 million galaxy redshifts up to $z=1.7$ and 
600,000 quasars ($2.2 < z < 3.5$) over 14,000 deg$^2$ \citep{Schelgel:2011zz}.
The proposed WFIRST\footnote{http://wfirst.gsfc.nasa.gov/} satellite would map 17 million galaxies in the redshift
range $1.3 < z < 2.7$ over 3400 deg$^2$, with a larger area 
possible with an extended mission \citep{Green:2012mj}.

Large-scale structure data from galaxy redshift surveys can be analyzed using either 
the power spectrum or the two-point correlation function. Although these two methods are 
Fourier transforms of one another, the analysis processes, the statistical uncertainties, and the systematics are quite 
different and the results cannot be converted using Fourier transform 
directly because of the finite size of the survey volume. 
The SDSS-II Luminous Red Galaxy (LRG) \citep{Eisenstein:2001cq} data have been analyzed, and the cosmological results delivered, using both the power spectrum 
(see, e.g., \citealt{Tegmark:2003uf,Hutsi:2005qv,Padmanabhan:2006ku,Blake:2006kv,Percival:2007yw,Percival:2009xn,Reid:2009xm,Montesano:2011bp}), 
and the correlation function method (see, e.g., 
\citealt{Eisenstein:2005su,Okumura:2007br,Cabre:2008sz,Martinez:2008iu,Sanchez:2009jq,Kazin:2009cj,Chuang:2010dv,Samushia:2011cs,Padmanabhan:2012hf,Xu:2012fw,Oka:2013cba,Hemantha:2013sea}). 
Similar analysis have been also applied on the SDSS-III BOSS galaxy sample 
\citep{Anderson:2012sa,Manera:2012sc,Nuza:2012mw,Reid:2012sw,Samushia:2012iq,Tojeiro:2012rp, Anderson:2013oza, Chuang:2013hya, Sanchez:2013uxa, Kazin:2013rxa, Wang:2014qoa, Anderson:2013zyy,Beutler:2013yhm,Samushia:2013yga,Tojeiro:2014eea,Reid:2014iaa,Alam:2015qta,Gil-Marin:2015nqa,Gil-Marin:2015sqa,Cuesta:2015mqa}.

Galaxy clustering allows us to differentiate smooth dark energy and modified gravity as the cause for cosmic acceleration through the simultaneous measurements 
of the cosmic expansion history $H(z)$ and the growth rate of cosmic large scale structure, $f(z)$ \citep{Guzzo08,Wang08,Blake:2012pj}. 
However, to measure $f(z)$, one must determine the galaxy bias $b$, which requires measuring higher-order statistics of the galaxy clustering (see \citealt{Verde:2001sf}).
\cite{Song09} proposed using the normalized growth rate, $f(z)\sigma_8(z)$, which would avoid the uncertainties from the galaxy bias.
\cite{Percival:2008sh} developed a method to measure $f(z)\sigma_8(z)$ and applied it on simulations.
\cite{Wang12} estimated expected statistical constraints on
dark energy and modified gravity, including redshift-space distortions and other constraints from galaxy clustering, using a Fisher matrix formalism.

In principle, the Hubble expansion rate $H(z)$, the angular-diameter distance $D_A(z)$, the normalized growth rate $f(z)\sigma_8(z)$, and
the physical matter density $\Omega_mh^2$ can be well constrained by analyzing the galaxy clustering data alone.
\cite{Eisenstein:2005su} demonstrated the feasibility of measuring $\Omega_mh^2$ and an effective distance, $D_V(z)$, 
from the SDSS DR3 \citep{Abazajian:2004it} LRGs, where $D_V(z)$ corresponds to a combination of $H(z)$ and $D_A(z)$. 
\cite{Chuang:2011fy} measured $H(z)$ and $D_A(z)$ simultaneously using the galaxy clustering data from 
the two dimensional two-point correlation function of SDSS DR7 \citep{Abazajian:2008wr} LRGs.
\cite{Chuang:2012ad,Chuang:2012qt} improved the method and modelling to measure $H(z)$, $D_A(z)$, $f(z)\sigma_8(z)$, and $\Omega_m h^2$ from the same data.

\cite{Samushia:2011cs} determined $f(z)\sigma_8(z)$ from the SDSS DR7 LRGs.
\cite{Blake:2012pj} measured $H(z)$, $D_A(z)$, and $f(z)\sigma_8(z)$ from the WiggleZ Dark Energy Survey galaxy sample.
\cite{Reid:2012sw} and \cite{Chuang:2013hya} measured $H(z)$, $D_A(z)$, and $f(z)\sigma_8(z)$ from the SDSS BOSS DR9 CMASS.

In this study, we apply the similar approach as \cite{Chuang:2012ad,Chuang:2012qt} and \cite{Chuang:2013hya} to determine $H(z)$, $D_A(z)$, and $f(z)\sigma_8(z)$, which extracts a summary of the cosmological information from the large-scale structure of the SDSS BOSS DR12 CMASS alone
by using very wide flat priors on the cosmological parameters which are not well constrained by galaxy clustering. 
We make some modifications from the methodologies used in previous works. First, we extract the cosmological information only using the correlation function at very large scales, i.e. $>55h^{-1}$Mpc to minimize the uncertainties from the effect at smaller scales, e.g., nonlinear effect, nonlinear redshift space distortion, and scale-dependent bias. Note that this strategy can only be applied to the analyses in configuration space since in Fourier space the uncertainties at small scales will propagate to wide $k$ range. We will validate our method using mock catalogues.
Second, it is known that some observational systematics can distort the observed galaxy clustering at the large scales we are interested in (e.g., \citealt{Ross:2012qm}). Although we apply the systematics weights to minimize their impact (see \citep{Reid:2015gra}), it is not granted that we have removed them completely. In this study, we include a polynomial as the model correcting observational systematic errors, e.g., ones induced by stars and seeing. We will show that our measurements are robust even in the case that we do not use the systematic weight corrections.
One can combine our single-probe measurements with other data sets (i.e. CMB, SNe, etc.) to constrain the cosmological parameters of a given dark energy model.

This paper is organized as follows. In Section \ref{sec:data}, we introduce the SDSS-III/BOSS DR12 CMASS galaxy sample and mock catalogues used 
in our study. In Section \ref{sec:method}, we describe the details of the 
methodology that constrains cosmological parameters from our galaxy clustering analysis. 
In Section \ref{sec:results}, we present our single-probe cosmological measurements. 
In Section \ref{sec:models}, given some simple dark energy models, we present the cosmological constraints from our measurements and the combination with other data sets.
In Section \ref{sec:compare}, we compare our measurements with the prediction of Planck assuming $\Lambda$CDM and other measurements obtained from galaxy clustering data.
We summarize and conclude in Section \ref{sec:conclusion}.
\section{Data sets} \label{sec:data}

\subsection{The CMASS Galaxy Catalogues}
\label{sec:cmass}

The Sloan Digital Sky Survey (SDSS; \citealt{Fukugita:1996qt,Gunn:1998vh,York:2000gk,Smee:2012wd}) mapped over one quarter 
of the sky using the dedicated 2.5 m Sloan Telescope \citep{Gunn:2006tw}.
The Baryon Oscillation Sky Survey (BOSS, \citealt{Eisenstein11, Bolton:2012hz, Dawson13}) is part of the SDSS-III survey. 
It is collecting the spectra and redshifts for 1.5 million galaxies, 160,000 quasars and 
100,000 ancillary targets. The Data Release 12 \citep{Alam:2015mbd} has been made publicly available\footnote{http://www.sdss3.org/}.
We use galaxies from the SDSS-III BOSS DR12 CMASS catalogue in the redshift range $0.43<z<0.75$.
CMASS samples are selected with an approximately constant stellar mass threshold \citep{Eisenstein11}; 
We are using 800853 CMASS galaxies.
The effective redshifts of the sample are $z=0.59$.
The details of generating this sample are described in \cite{Reid:2015gra}.

\subsection{The Mock Catalogues}

For the data release 9, 10, and 11, PTHalos mock catalogues \cite{Manera:2012sc, Manera13} were used for constructing the covariance matrix of the clustering measurements. 
For the data release 12 (this study), we use 2000 BOSS DR12 MultiDark-PATCHY (MD-PATCHY) mock galaxy catalogues \citep{Kitaura:2015uqa} for validating our methodology and estimating the covariance matrix in this study. These mock catalogues were constructed using a similar procedure described in \citealt{Rodriguez-Torres:2015vqa} where they constructed the BOSS DR12 lightcone mock catalogues using the MultiDark $N$-body simulations \citep{Klypin:2014kpa}. However, instead of using $N$-body simulations, the 2000 MD-PATCHY mocks catalogues were constructed using the PATCHY approximate simulations.
These mocks are produced using ten boxes at different redshifts that are created with the PATCHY-code \citep{Kitaura:2013cwa}. The PATCHY-code can be composed into two parts: 1) computing approximate dark matter density field; and 2) populating galaxies from dark matter density field with the biasing model. The dark matter density field is estimated using Augmented Lagrangian Perturbation Theory (ALPT; \cite{Kitaura:2012tj}) which combines the second order perturbation theory (2LPT) and spherical collapse approximation. The biasing model includes deterministic bias and stochastic bias (see \cite{Kitaura:2013cwa,Kitaura:2014mja} for details). The velocity field is constructed based on the displacement field of dark matter particles. The modeling of finger-of-god has also been taken into account statistically. The mocks match the clustering of the galaxy catalogues for each redshift bin (see \cite{Kitaura:2015uqa} for details).
The mock catalogues were constructed assuming $\Lambda$CDM Planck cosmology with \{$\Omega_{\rm M}=0.307115, \Omega_{\rm b}=0.048206,\sigma_8=0.8288,n_s=0.96$\}, and a Hubble constant ($H_0=100\,h\Hunit$) given by  $h=0.6777$. 
As shown in a mock catalogue comparison study (\citealt{Chuang:2014toa}), PATCHY mocks are accurate within 5\% on scales larger than 5 Mpc/h (or $k$ smaller than 0.5 h/Mpc in Fourier space) for monopole and within 10-15\% for quadrupole. \cite{Kitaura:2015uqa} had also demonstrated the accuracy of BOSS PATCHY mock catalogues which are in very good agreement with the observed data in terms of 2- and 3-point statistics.

\section{Methodology} 
\label{sec:method}

In this section, we describe the measurement of the multipoles of the correlation function
from the observational data, construction of the theoretical prediction, 
and the likelihood analysis that leads to constraining 
cosmological parameters and dark energy. 

\subsection{Measuring the Two-Dimensional Two-Point Correlation Function}

We convert the measured redshifts of the BOSS CMASS galaxies to comoving distances 
by assuming a fiducial model, i.e., flat $\Lambda$CDM with $\Omega_m=0.307115$ and $h=0.6777$ 
which is the same model adopted for constructing the mock catalogues (see \citealt{Kitaura:2015uqa}). 
We use the two-point correlation function estimator given by 
\cite{Landy:1993yu}:
\begin{equation}
\label{eq:xi_Landy}
\xi(s,\mu) = \frac{DD(s,\mu)-2DR(s,\mu)+RR(s,\mu)}{RR(s,\mu)},
\end{equation}
where $s$ is the separation of a pair of objects and $\mu$ is the cosine of the angle between the directions between the line of sight (LOS) and the line connecting the pair the objects. DD, DR, and RR represent the normalized data-data,
data-random, and random-random pair counts, respectively, for a given
distance range. The LOS is defined as the direction from the observer to the 
centre of a galaxy pair. Our bin size is
$\Delta s=1 \, h^{-1}$Mpc and $\Delta \mu=0.01$. 
The Landy and Szalay estimator has minimal variance for a Poisson
process. Random data are generated with the same radial
and angular selection functions as the real data. One can reduce the shot noise due
to random data by increasing the amount of random data. The number
of random data we use is about 50 times that of the real data. While
calculating the pair counts, we assign to each data point a radial
weight of $1/[1+n(z)\cdot P_w]$, where $n(z)$ is the radial
number density and $P_w = 10^4$ $h^{-3}$Mpc$^3$ (see  
\citealt{Feldman:1993ky}).
We include the combination of the observational weights assigned for each galaxy by
\begin{equation}\label{eq:weight}
w_{tot,i} = w_{sys,i}*(w_{rf,i}+w_{fc,i}-1),
\end{equation}
where $w_{tot,i}$ is the final weight to assign on a galaxy $i$; $w_{sys,i}$ is for removing the correlation between CMASS galaxies and both stellar density and seeing; $w_{rf,i}$ and $w_{fc,i}$ correct for missing objects due to the redshift failure and fiber collision. The details are described in \cite{Reid:2015gra} (see also \citealt{Ross:2012qm}). Later, we will also test the impact of systematics by removing $w_{sys,i}$ from the analysis.

\subsection{Theoretical Two-Dimensional Two-Point Correlation Function}
\label{sec:model_large}

The theoretical model for linear and quasi-linear scales can be constructed by first and higher order perturbation theory. 
One can compute the model by adding the first order nonlinear corrections to the linear theoretical model. 
There is no other fitting parameter besides the cosmological parameters (which will be introduced later in this paper).
The procedure of constructing theoretical model for quasi-linear scales in redshift space is the following:
First, we adopt the cold dark matter model and the simplest inflation model (adiabatic initial condition).
Thus, we can compute the linear matter power spectra, $P_{lin}(k)$, by using CAMB (Code for Anisotropies in the Microwave Background, \citealt{Lewis:1999bs}). The linear power spectrum can be decomposed into two parts:
\begin{equation} \label{eq:pk_lin}
P_{lin}(k)=P_{nw}(k)+P_{BAO}^{lin}(k),
\end{equation}
where $P_{nw}(k)$ is the ``no-wiggle'' or pure CDM power spectrum calculated using Eq.(29) from \cite{Eisenstein:1997ik}. $P_{BAO}^{lin}(k)$ is the ``wiggled'' part defined by Eq. (\ref{eq:pk_lin}).
The nonlinear damping effect of the ``wiggled'' part, in redshift space, can be well approximated following \cite{Eisenstein:2006nj} by
\begin{equation} \label{eq:bao}
P_{BAO}^{nl}(k,\mu_k)=P_{BAO}^{lin}(k)\cdot \exp\left(-\frac{k^2}{2k_\star^2}[1+\mu_k^2(2f+f^2)]\right),
\end{equation}
where $\mu_k$ is the cosine of the angle between ${\bf k}$ and the LOS, $f$ is the growth rate, and
$k_\star$ is computed following \cite{Crocce:2005xz} and \cite{Matsubara:2007wj} by
\begin{equation} \label{eq:kstar}
k_\star=\left[\frac{1}{3\pi^2}\int P_{lin}(k)dk\right]^{-1/2}.
\end{equation}
The dewiggled power spectrum is
\begin{equation} \label{eq:pk_dw}
P_{dw}(k,\mu_k)=P_{nw}(k)+P_{BAO}^{nl}(k,\mu_k).
\end{equation}

Besides the nonlinear redshift distortion introduced above, we include the linear redshift distortion as follows in order to
obtain the galaxy power spectrum in redshift space at large scales \citep{Kaiser:1987qv},
\begin{eqnarray} \label{eq:pk_2d}
P_g^s(k,\mu_k)&=&b^2(1+\beta\mu_k^2)^2P_{dw}(k,\mu_k),
\end{eqnarray}
where $b$ is the linear galaxy bias and $\beta$ is the linear redshift distortion parameter.

We compute the theoretical two-point correlation 
function, $\xi_{th}(\sigma,\pi)$, for quasi-linear scales by Fourier transforming the non-linear power spectrum
$P_g^s(k,\mu_k)$. This task is efficiently performed by using Legendre polynomial expansions and one-dimensional integral convolutions as introduced in \cite{Chuang:2012qt}.
Power spectrum analysis is more sensitive to the nonlinear effects than the correlation function analysis since the uncertainty at small scales would propagate to wider range of k. To have some idea, one can compare Fig 4 and Fig 7 in \cite{Chuang:2014toa} and will find that different mock catalogues have similar performance in configuration space but are very different in k-space. As shown in the \cite{Eisenstein:2006nj}, the damping of BAO is the major correction of the nonlinear effects in the configuration space at the scales interested, e.g. $s>55\,h^{-1}$Mpc.
In Fig 7 of \cite{Samushia:2013yga}, they showed that the growth rate measured using linear redshift distortion model could be biased by 3\% when using the scales larger than $55 \,h^{-1}$Mpc. The accuracy is acceptable since the uncertainty of our $f(z)\sigma_8(z)$ measurement is about 12\%.

\subsection{Measure Multipoles of the Two-Point Correlation Function}  \label{sec:multipoles}

The traditional multipoles of the two-point correlation function, in redshift space, are defined by
\ba
\label{eq:multipole_1}
\xi_l(s) &\equiv & \frac{2l+1}{2}\int_{-1}^{1}{\rm d}\mu\, \xi(s,\mu) P_l(\mu),
\ea
where $P_l(\mu)$ is the Legendre Polynomial ($l=$0 and 2 here). 
We integrate over a spherical shell with radius $s$,
while actual measurements of $\xi(s,\mu)$ are done in discrete bins.
To compare the measured $\xi(s,\mu)$ and our theoretical model, the last integral in Eq.(\ref{eq:multipole_1}) should be converted into a sum,
\begin{equation}\label{eq:multipole}
 \hat{\xi}_l(s) \equiv \frac{\displaystyle\sum_{s-\frac{\Delta s}{2} < s' < s+\frac{\Delta s}{2}}\displaystyle\sum_{0\leq\mu\leq1}(2l+1)\xi(s',\mu)P_l(\mu)}{\mbox{Number of bins used in the numerator}},
\end{equation}
where $\Delta s=5$ $h^{-1}$Mpc in this work.

We are using the scale range $s=55-200\,h^{-1}$Mpc and the bin size is 5 $h^{-1}$Mpc. 
The data points from the multipoles in the scale range considered are combined to form a 
vector, $X$, i.e.,
\be
{\bf X}=\{\hat{\xi}_0^{(1)}, \hat{\xi}_0^{(2)}, ..., \hat{\xi}_0^{(N)}; 
\hat{\xi}_2^{(1)}, \hat{\xi}_2^{(2)}, ..., \hat{\xi}_2^{(N)};...\},
\label{eq:X}
\ee
where $N$ is the number of data points in each measured multipole; here $N=29$.
The length of the data vector ${\bf X}$ depends on the number of multipoles used.

\subsection{Model for Systematic Errors}

It is well known that the observations could be contaminated by systematic effects. 
To obtain the robust and conservative measurements, we include a model for systematics.
The model is a simple polynomial given by
\begin{equation} \label{eq:systematic}
 A(s)=a_0+\frac{a_1}{s}+\frac{a_2}{s^2}.
\end{equation}
Since the quadrupole is insensitive to the systematics effects of which we are aware (see Fig. \ref{fig:multipoles_cmass} or more details in \citealt{Ross:2012qm}), we include the systematics model for only the monopole of the theoretical model by
\begin{equation}
\xi’_{0,th}(s) = \xi_{0,th}(s) + A(s),
\end{equation}
where $\xi_{0,th}(s)$ is the monopole derived from $\xi_{th}(\sigma,\pi)$ in Sec. \ref{sec:model_large}.
Note that $A(s)$ is in the same form as the smooth function used in the BAO-only analyses (e.g., see \citealt{Xu:2012fw,Anderson:2013zyy}). In those analyses,  two smooth functions have been applied to remove the full shape information of monopole and quadrupole respectively. However, if we added the smooth function to quadrupole, we would not be able to measure $f(z)\sigma_8(z)$. Fortunately, the quadrupole is insensitive to the systematics as shown in Fig. \ref{fig:multipoles_cmass}, so that we do not remove its full shape information and thus we can measure $f(z)\sigma_8(z)$.

\subsection{Covariance Matrix} \label{sec:covar}

We use the 2000 mock catalogues created by \citealt{Kitaura:2015uqa}
for the BOSS DR12 CMASS galaxy sample
to estimate the covariance matrix of the observed correlation function. 
We calculate the multipoles of the correlation functions 
of the mock catalogues and construct the covariance matrix as
\begin{equation}
 C_{ij}=\frac{1}{(N-1)(1-D)}\sum^N_{k=1}(\bar{X}_i-X_i^k)(\bar{X}_j-X_j^k),
\label{eq:covmat}
\end{equation}
where
\begin{equation}
 D=\frac{N_b +1}{N-1},
\label{eq:D}
\end{equation}
$N$ is the number of the mock catalogues, $N_b$ is the number of data bins, $\bar{X}_m$ is the
mean of the $m^{th}$ element of the vector from the mock catalogue multipoles, and
$X_m^k$ is the value in the $m^{th}$ elements of the vector from the $k^{th}$ mock
catalogue multipoles. The data vector ${\bf X}$ is defined by Eq.(\ref{eq:X}).
We also include the correction, $D$, introduced by \cite{Hartlap:2006kj}. 

\subsection{Likelihood}
The likelihood is taken to be proportional to $\exp(-\chi^2/2)$ \citep{press92}, 
with $\chi^2$ given by
\begin{equation} \label{eq:chi2}
 \chi^2\equiv\sum_{i,j=1}^{N_{X}}\left[X_{th,i}-X_{obs,i}\right]
 C_{ij}^{-1}
 \left[X_{th,j}-X_{obs,j}\right]
\end{equation}
where $N_{X}$ is the length of the vector used, 
$X_{th}$ is the vector from the theoretical model, and $X_{obs}$ 
is the vector from the observed data.

As explained in \cite{Chuang:2011fy}, instead of recalculating the observed correlation function while 
computing for different models, we rescale the theoretical correlation function to avoid rendering the $\chi^2$ values arbitrary.
This approach can be considered as an application of Alcock-Paczynski effect \citep{Alcock:1979mp}.
The rescaled theoretical correlation function is computed by
\begin{equation} \label{eq:inverse_theory_2d}
 T^{-1}(\xi_{th}(\sigma,\pi))=\xi_{th}
 \left(\frac{D_A(z)}{D_A^{fid}(z)}\sigma,
 \frac{H^{fid}(z)}{H(z)}\pi\right),
\end{equation}
where $\xi_{th}$ is defined in Sec. \ref{sec:model_large} and $\chi^2$ can be rewritten as
\ba 
\label{eq:chi2_2}
\chi^2 &\equiv&\sum_{i,j=1}^{N_{X}}
 \left\{T^{-1}X_{th,i}-X^{fid}_{obs,i}\right\}
 C_{fid,ij}^{-1} \cdot \nonumber\\
 & & \cdot \left\{T^{-1}X_{th,j}-X_{obs,j}^{fid}\right\};
\ea
where $T^{-1}X_{th}$ is the vector computed by eq.\ (\ref{eq:multipole}) from the rescaled theoretical correlation function, eq. (\ref{eq:inverse_theory_2d}), taking into account the modeling of observational systematics, eq. (\ref{eq:systematic}).
$X^{fid}_{obs}$ is the vector from observed data measured with the fiducial model (see \citealt{Chuang:2011fy} for more details regarding the rescaling method).

\subsection{Markov Chain Monte-Carlo Likelihood Analysis} \label{sec:mcmc}

We perform Markov Chain Monte-Carlo
likelihood analyses using CosmoMC \citep{Lewis:2002ah}. 
The parameter space that we explore spans the parameter set of
$\{H(z)$, $D_A(z)$, $\Omega_mh^2$, $\beta(z)$, $b\sigma_8(z)$, $\Omega_bh^2$, $n_s$, $f(z)$, $a_0$, $a_1$, $a_2$$\}$. 
The quantities $\Omega_m$ and $\Omega_b$ are the matter and
baryon density fractions, $n_s$ is the power-law index of the primordial matter power spectrum, 
$h$ is the dimensionless Hubble
constant ($H_0=100h$ km s$^{-1}$Mpc$^{-1}$), and $\sigma_8(z)$ is the normalization of the power spectrum.
The linear redshift distortion parameter can be expressed as $\beta(z)=f(z)/b$.
Thus, one can derive $f(z)\sigma_8(z)$ from the measured $\beta(z)$ and $b\sigma_8(z)$.
Among these parameters, only $\{H(z)$, $D_A(z)$, $\Omega_mh^2$, $\beta(z)$, $b\sigma_8(z)\}$ are well constrained using
the BOSS galaxy sample alone in the scale range of interest. We marginalize over the other parameters, 
$\{\Omega_bh^2$, $n_s$, $f(0.59)$, $a_0$, $a_1$, $a_2\}$, with the flat priors 
$\{(0.018768, 0.025368)$, $(0.8684, 1.0564)$, $(0.3, 1)$, $(-0.003, 0.003)$, $(-3, 3)$, $(-20, 20)\}$, 
where the flat priors of $\Omega_b h^2$ and $n_s$ are centered on 
the Planck measurements with a width of $\pm10\sigma_{Planck}$ ($\sigma_{Planck}$ is taken from \citealt{Ade:2013zuv}). These priors
are sufficiently wide to ensure that CMB constraints are not double counted 
when our results are combined with CMB data \citep{Chuang:2010dv}.

On the scales we use for comparison with the BOSS galaxy data, the theoretical correlation 
function only depends on cosmic curvature and dark energy through the
parameters $H(z)$, $D_A(z)$, $\beta(z)$, and $b\sigma_8(z)$        
assuming that dark energy perturbations are unimportant (valid in the simplest dark energy models).
Thus we are able to extract constraints from clustering data that are independent of dark energy.

\section{Results} \label{sec:results}
\begin{figure*}
\begin{center}
 \subfigure{\label{fig:mono}\includegraphics[width=0.9 \columnwidth,clip,angle=0]{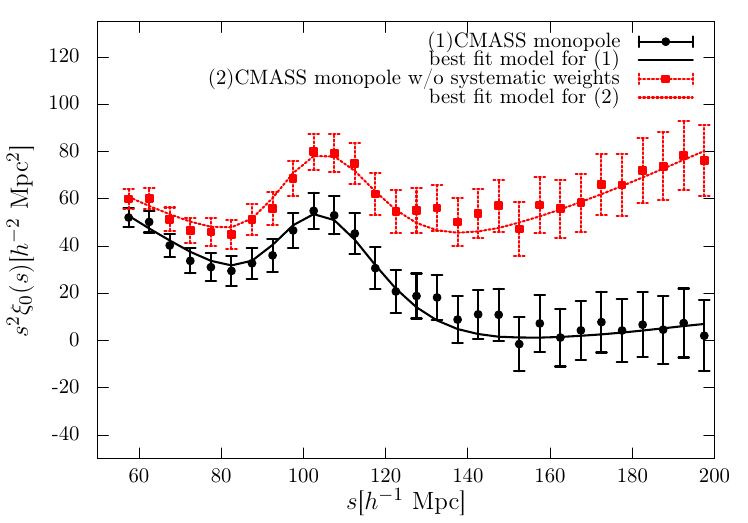}}
 \subfigure{\label{fig:quad}\includegraphics[width=0.9 \columnwidth,clip,angle=0]{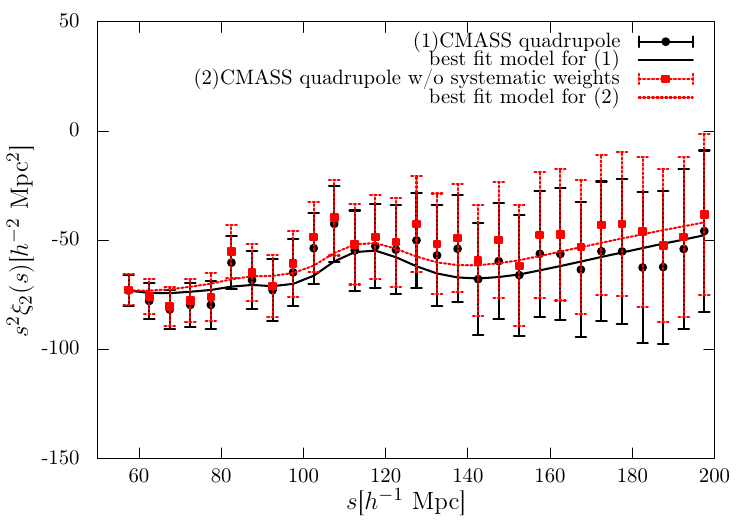}}
\end{center}
\caption{
Measurement of effective monopole (left) and quadrupole (right) of the correlation function from the BOSS DR12 CMASS galaxy sample with/without systematics weights for star and seeing(black/red points), compared to the theoretical models 
given the parameters measured (solid lines).
The error bars are the square roots of the diagonal elements of the covariance matrix.
In this study, our fitting scale ranges are $55h^{-1}$Mpc $<s<200h^{-1}$Mpc; the bin size is $5h^{-1}$Mpc.
The minimum $\chi^2$ per degree of freedom is 0.95 for the correlation function computed including the systematics weights; the one without including the systematics weights is 1.05.
}
\label{fig:multipoles_cmass}
\end{figure*}

Fig.\ref{fig:multipoles_cmass} shows the effective 
monopole ($\hat{\xi}_0$) and quadrupole ($\hat{\xi}_2$) measured from the BOSS CMASS galaxy sample
compared with the theoretical models given the parameters measured.
For the CMASS sample, we also present the correlation function measured from the sample without including systematics weights for stars and seeing. We do not test with the systematics weights for fiber collisions and redshift failures because those only affect smallest scales (i.e. $s<20h^{-1}$Mpc, see \citealt{Ross:2012qm}).
We will show that the measurements from our methodology are robust against these systematics. The minimum $\chi^2$ per degree of freedom is 0.95 for the correlation function computed including the systematics weights; the one without including the systematics weights is 1.05.

\subsection{Measurements of Cosmological Parameters}
\label{sec:CMASS_result}

With the increasing volume of the galaxy survey, one can obtain the cosmological constraints 
using the scales which can be modelled simply by perturbation theory (see Sec. \ref{sec:model_large}).
We now present the dark energy model independent measurements of the parameters
$\{H(0.59)$, $D_A(0.59)$, $\Omega_m h^2$, $\beta(0.59)$, and $b\sigma_8(0.59)\}$, obtained by using the 
method described in previous sections. We also present the derived 
parameters including $H^{-1}(0.59)r_s/r_{s,fid}$, $D_A(0.59)r_{s,fid}/r_s$, and 
$D_V(0.59)r_{s,fid}/r_s$ with
\begin{equation} \label{eq:dv}
 D_V(z)\equiv \left[(1+z)^2D_A(z)^2\frac{cz}{H(z)}\right]^\frac{1}{3},
\end{equation} 

where $r_s$ is the comoving sound horizon at the drag epoch calculated
using eq. (6) by CAMB
and $r_{s,fid}=148.66$Mpc is the $r_s$ of the fiducial cosmology used in this study (same as the one used by the mock catalogues).
$D_V(z)$ is the effective distance which can be measured from the spherical averaged correlation function or power spectrum (e.g. see \citealt{Eisenstein:2005su}).

While $H(z)r_s/r_{s,fid}$ and $D_A(z)r_{s,fid}/r_s$ measurements are mainly determined by the BAO feature, $\Omega_m h^2$ is basically determined by the overall shape. In table \ref{table:cov_cmass}, one can see  that the correlations between $\Omega_m h^2$ and both $H(z)r_s/r_{s,fid}$ and $D_A(z)r_{s,fid}/r_s$ are small.
Since the measurement of monopole is sensitive to the systematics (see Fig. \ref{fig:multipoles_cmass}), the measurement of $\Omega_m h^2$ would be also sensitive to the systematics and the constraint is weak. However, we still include $\Omega_mh^2$ while using our results to take into account the correlations between $f(z)\sigma_8(z)$ and $\Omega_m h^2$.

Table \ref{table:patchy_test} present our test using the mock catalogues. We apply our methodology on the mean of 2000 correlation functions from the mock catalogues and restore their input values within 0.4 $\sigma$ which shows that one can obtain reasonable results even with such simple model we are using. Note that the simplicity/speed of the model is critical for this work since we are scanning very large parameter space (with wide flat priors) and including the nuisance parameters for modeling the observational systematics. We will investigate more accurate models in the future work (Chuang et al. in prep.)

Table \ref{table:cmass} lists the mean, rms variance, and 68\%
confidence level limits for $H^{-1}(0.59)r_s/r_{s,fid}$, $D_A(0.59)r_{s,fid}/r_s$, $f(0.59)\sigma_8(0.59)$, $\Omega_mh^2$, and 
$D_V(0.59)r_{s,fid}/r_s\}$ derived in an MCMC likelihood analysis from the measured $\hat{\xi}_0+\hat{\xi}_2$ 
of the DR12 CMASS correlation function. 

Table \ref{table:cov_cmass} presents the normalized covariance matrix
for this parameter set measured using $\hat{\xi}_0+\hat{\xi}_2$.
The correlation between $\Omega_m h^2$ and $H^{-1}(0.59)r_s/r_{s,fid}$ or $D_A(0.59)r_{s,fid}/r_s$ are close to zero. However, the correlation coefficient of $f(0.59)\sigma_8(0.59)$ and $\Omega_m h^2$ is about -0.5. Therefore, we include $\Omega_m h^2$ to our product while the constraint of $\Omega_m h^2$ is weak comparing to the one from CMB.

\begin{table}
\begin{center}
\begin{tabular}{crrr}
\hline
		&	mean of mocks			&	Input values	&	deviation		\\	\hline
$	D_A(0.59)r_{s,fid}/r_s	$&$	1417	\pm	28	$&$	1409.26	$&$	0.29	\sigma	$\\	
$	H(0.59)r_s/r_{s,fid}	$&$	94.4	\pm	3.6	$&$	94.09	$&$	0.09	\sigma	$\\	
$	f\sigma_8(0.59)	$&$	0.502	\pm	0.061	$&$	0.4786	$&$	0.38	\sigma	$\\	
$	\Omega_mh^2	$&$	0.144	\pm	0.016	$&$	0.14105	$&$	0.20	\sigma	$\\	
$	D_V(0.59)r_{s,fid}/r_s	$&$	2120	\pm	31	$&$	2113.37	$&$	0.20	\sigma	$\\	
\hline
\end{tabular}
\end{center}
\caption{Test using the mean of the correlation functions from the mock catalogues.
We restore the input values within $0.4\sigma$.
The units of $H$ are $\Hunit$, the units of $D_A$ and $D_V$ are $\rm Mpc$, and $\omega_m$ is defined as $\Omega_m h^2$. 
} \label{table:patchy_test}
\end{table}

\begin{table}
\begin{center}
\begin{tabular}{crrr}
\hline
		&	fiducial results			&	no sys. weights			&	difference		\\	\hline
$	D_A(0.59)r_{s,fid}/r_s	$&$	1427	\pm	26	$&$	1422	\pm	27	$&$	0.18	\sigma	$\\	
$	H(0.59)r_s/r_{s,fid}	$&$	97.3	\pm	3.3	$&$	96.7	\pm	3.4	$&$	0.19	\sigma	$\\	
$	f\sigma_8(0.59)	$&$	0.488	\pm	0.060	$&$	0.479	\pm	0.060	$&$	0.15	\sigma	$\\	
$	\Omega_mh^2	$&$	0.135	\pm	0.016	$&$	0.137	\pm	0.016	$&$	0.12	\sigma	$\\	
$	D_V(0.59)r_{s,fid}/r_s	$&$	2107	\pm	27	$&$	2107	\pm	28	$&$	0.01	\sigma	$\\	
\hline
\end{tabular}
\end{center}
\caption{The fiducial measurement and systematic test from the correlation function of DR12 CMASS sample. The systematics test is using the observed correlation function without including the systematics weights (i.e. star and seeing).
One can see the measured quantities are robust against these systematics.
The units of $H$ are $\Hunit$, the units of $D_A$ and $D_V$ are $\rm Mpc$.
} \label{table:cmass}
\end{table}

\begin{table*}
\begin{center}
\begin{tabular}{crrrrr}
\hline
\\
		&$	\frac{D_A(0.59)}{r_s/r_{s,fid}}	$&$	\frac{H(0.59)}{r_{s,fid}/r_s}	$&$	f\sigma_8(0.59)	$&$	\Omega_mh^2	$&$	\frac{D_V(0.59)}{r_s/r_{s,fid}}	$\\	\hline
$	D_A(0.59)r_{s,fid}/r_s	$&$	1.0000	$&$	0.4129	$&$	0.2806	$&$	0.1266	$&$	0.5849	$\\	
$	H(0.59)r_s/r_{s,fid}	$&$	0.4129	$&$	1.0000	$&$	0.2897	$&$	0.0543	$&$	-0.4969	$\\	
$	f\sigma_8(0.59)	$&$	0.2806	$&$	0.2897	$&$	1.0000	$&$	-0.4856	$&$	0.0091	$\\	
$	\Omega_mh^2	$&$	0.1266	$&$	0.0543	$&$	-0.4856	$&$	1.0000	$&$	0.0742	$\\	
$	D_V(0.59)r_{s,fid}/r_s	$&$	0.5849	$&$	-0.4969	$&$	0.0091	$&$	0.0742	$&$	1.0000	$\\	
\hline
\end{tabular}

\end{center}
\caption{Normalized covariance matrix of the fiducial measurements from CMASS galaxy sample (using $55<s<200\ h^{-1}$Mpc).
The units of $H$ are $\Hunit$, the units of $D_A$ and $D_V$ are $\rm Mpc$.
} \label{table:cov_cmass}
\end{table*}

\subsection{Using Our Results from Galaxy Clustering only}
In this section, we describe the steps to combine our results with other data sets assuming some dark energy models. 
Here, we use the results from CMASS quasi-linear scales as an example.
For a given model and cosmological parameters, one can compute 
$H^{-1}(0.59)r_s/r_{s,fid}$, $D_A(0.59)r_{s,fid}/r_s$, $f(0.59)\sigma_8(0.59)$, $\Omega_mh^2$. From Table \ref{table:cmass} and \ref{table:cov_cmass},
one can derive the covariance matrix, $M_{ij}$, of these three parameters. Then, $\chi^2$ can be computed by
\begin{equation}
 \chi^2=\Delta_{CMASS}M_{ij}^{-1}\Delta_{CMASS},
\end{equation}
where 
\begingroup
\everymath{\scriptstyle}
\small
\begin{equation}
 \Delta_{CMASS}=\left(\begin{array}{c}
D_A(0.59)r_{s,fid}/r_s-1427 \\ 
H(0.59)r_s/r_{s,fid}-97.3 \\ 
f(0.59)\sigma_8(0.59)-0.488 \\
\Omega_m h^2 - 0.135
\end{array}\right)
\end{equation}
\endgroup
and
\begingroup
\everymath{\scriptstyle}
\small
\begin{equation}
M_{ij}=\left(\begin{array}{cccc}
6.77E+02	&	3.58E+01	&	4.36E-01	&	5.24E-02	\\
3.58E+01	&	1.11E+01	&	5.77E-02	&	2.88E-03	\\
4.36E-01	&	5.77E-02	&	3.57E-03	&	-4.62E-04	\\
5.24E-02	&	2.88E-03	&	-4.62E-04	&	2.53E-04	
\end{array}\right),
\end{equation}
\endgroup
where $M_{ij}$ can be derived from table \ref{table:cmass} and \ref{table:cov_cmass}. Note table \ref{table:cov_cmass} shows the normalized covariance matrix $N_{ij}$, and $M_{ij}$ can be derived by
$M_{ij} = N_{ij}  \sigma_i  \sigma_j$,
where $\sigma_i$ or $\sigma_j$ are the standard deviations of the fiducial results in table \ref{table:cmass}.

\section{Assuming Dark Energy Models} 
\label{sec:models}
In this section, we present examples of combining our CMASS clustering results with the Planck CMB data \citep{Adam:2015rua} assuming specific dark energy models. 

\begin{figure}
\centering
\includegraphics[width=1 \columnwidth,clip,angle=0]{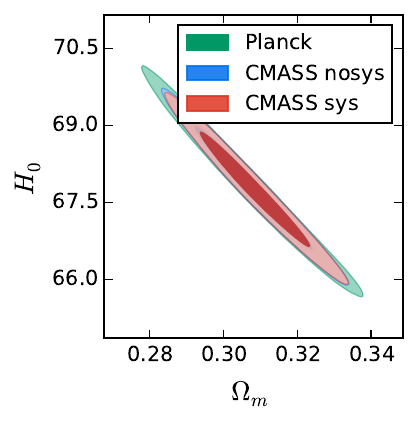}
\caption{
2D marginalized
  contours for $68\%$ and $95\%$ confidence levels for $\Omega_m$ and $H_0$ ($\Lambda$CDM model assumed)
from Planck-only (green), Planck+CMASS  (red), and Planck+CMASS without including systematics weights (blue).
}
\label{fig:om_h}
\end{figure}

\begin{figure}
\centering
\includegraphics[width=1 \columnwidth,clip,angle=0]{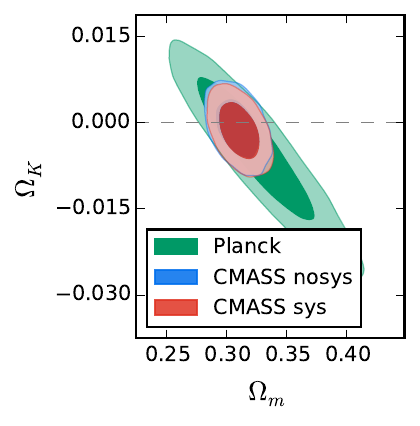}
\caption{
2D marginalized
  contours for $68\%$ and $95\%$ confidence level for $\Omega_m$ and $\Omega_k$ (o$\Lambda$CDM model assumed)
from Planck-only (green), Planck+CMASS  (red), and Planck+CMASS without including systematics weights (blue).
One can see that $\Omega_k$ is consistent with 0 which is corresponding to the flat universe.
}
\label{fig:om_ok}
\end{figure}

\begin{figure}
\centering
\includegraphics[width=1 \columnwidth,clip,angle=0]{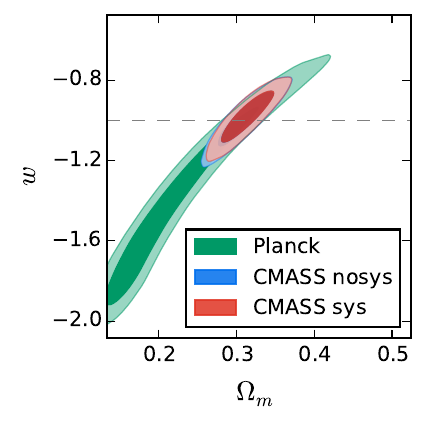}
\caption{
2D marginalized
  contours for $68\%$ and $95\%$ confidence level for $\Omega_m$ and $w$ ($w$CDM model assumed)
from Planck-only (green), Planck+CMASS  (red), and Planck+CMASS without including systematics weights (blue).
One can see that $w$ is consistent with -1 which is corresponding to the $\Lambda$CDM.
}
\label{fig:om_w}
\end{figure}

\begin{figure}
\centering
\includegraphics[width=1 \columnwidth,clip,angle=0]{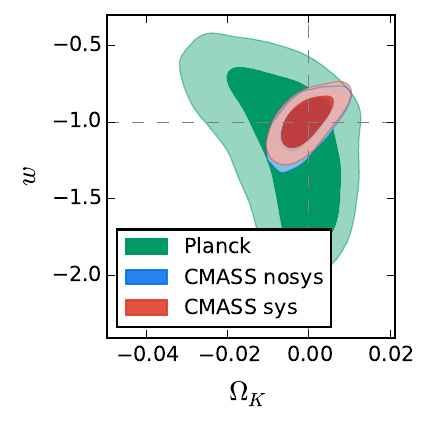}
\caption{
2D marginalized
  contours for $68\%$ and $95\%$ confidence level for $\Omega_k$ and $w$ ($w$CDM model assumed)
from Planck-only (green), Planck+CMASS  (red), and Planck+CMASS without including systematics weights (blue).
One can see that $\Omega_k$ is consistent with 0 and $w$ is consistent with -1 which is corresponding to the $\Lambda$CDM.
}
\label{fig:ok_w}
\end{figure}

\begin{figure}
\centering
\includegraphics[width=1 \columnwidth,clip,angle=0]{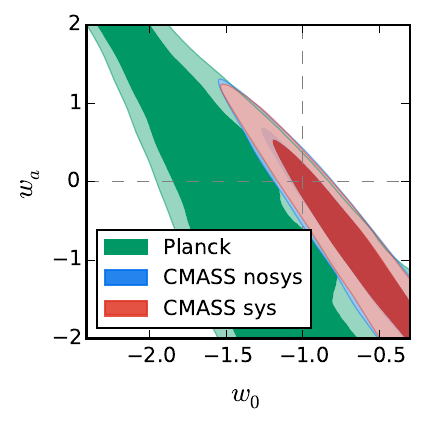}
\caption{
2D marginalized
  contours for $68\%$ and $95\%$ confidence level for $w_0$ and $w_a$ ($w_0w_a$CDM model assumed)
from Planck-only (green), Planck+CMASS  (red), and Planck+CMASS without including systematics weights (blue).
One can see that $w_0$ and $w_a$ are consistent with -1 and 0 respectively which are corresponding to the $\Lambda$CDM.
}
\label{fig:w0wa}
\end{figure}

\begin{figure}
\centering
\includegraphics[width=1 \columnwidth,clip,angle=0]{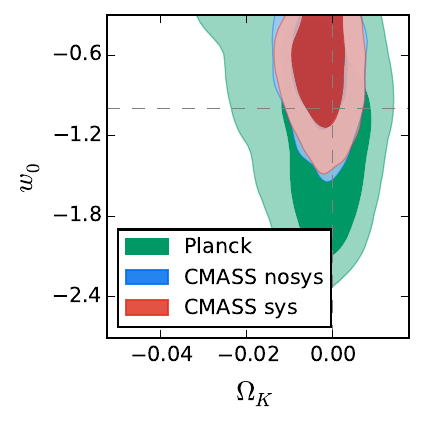}
\caption{
2D marginalized
  contours for $68\%$ and $95\%$ confidence level for $\Omega_k$ and $w_0$ (o$w_0w_a$CDM model assumed)
from Planck-only (green), Planck+CMASS  (red), and Planck+CMASS without including systematics weights (blue).
One can see that $\Omega_k$ and $w_0$ are consistent with 0 and -1 respectively which are corresponding to the $\Lambda$CDM.
}
\label{fig:ow0wa}
\end{figure}

Table \ref{table:DEmodels} shows the cosmological constraints assuming 
$\Lambda$CDM, o$\Lambda$CDM (non-flat $\Lambda$CDM), $w$CDM (constant equation of state of dark energy), o$w$CDM, $w_0w_a$CDM, and o$w_0w_a$CDM. 
Table \ref{table:DEmodels_nosys} shows the cosmological constraints obtained from the correlation function without observational systematics corrections. We find it agrees very well with Table \ref{table:DEmodels}.
We also present the 2D marginalized
  contours comparing with Planck CMB data \citep{Adam:2015rua} in Fig. \ref{fig:om_h}, \ref{fig:om_ok}, \ref{fig:om_w}, \ref{fig:ok_w}, \ref{fig:w0wa}, and \ref{fig:ow0wa}.
One can see that the constraints obtained from our measurements without including observational systematics weights agree very well with the corrected ones. In addition, we do not find any deviation from $\Lambda$CDM by testing various models.

\begin{table*}
\begin{center}
\begin{tabular}{lrrrrrr}
\hline
	&	$\Omega_m$	&			$H_0$	&			$\sigma_8$	&			$\Omega_k$	&			$w$ or $w_0$	&			$w_a$			\\	\hline
$\Lambda$CDM	& $	0.308	\pm	0.010	$&$	67.8	\pm	0.7	$&$	0.815	\pm	0.009	$&$	0			$&$	-1			$&$	0			$\\	
o$\Lambda$CDM	& $	0.311	\pm	0.011	$&$	67.4	\pm	1.1	$&$	0.812	\pm	0.010	$&$	-0.001	\pm	0.003	$&$	-1			$&$	0			$\\	
$w$CDM	& $	0.314	\pm	0.021	$&$	67.2	\pm	2.3	$&$	0.810	\pm	0.022	$&$	0			$&$	-0.98	\pm	0.08	$&$	0			$\\	
o$w$CDM	& $	0.313	\pm	0.024	$&$	67.3	\pm	2.4	$&$	0.810	\pm	0.024	$&$	-0.001	\pm	0.004	$&$	-0.99	\pm	0.11	$&$	0			$\\	
$w_0w_a$CDM	& $	0.332	\pm	0.032	$&$	65.5	\pm	3.2	$&$	0.796	\pm	0.028	$&$	0			$&$	-0.76	\pm	0.28	$&$	-0.63	\pm	0.73	$\\	
o$w_0w_a$CDM	& $	0.333	\pm	0.032	$&$	65.3	\pm	3.1	$&$	0.796	\pm	0.028	$&$	-0.003	\pm	0.004	$&$	-0.74	\pm	0.27	$&$	-0.80	\pm	0.74	$\\	
\hline
\end{tabular}
\end{center}
\caption{ 
The cosmological constraints from our CMASS measurements combining with Planck data assuming $\Lambda$CDM, nonflat $\Lambda$CDM (o$\Lambda$CDM), $w$CDM, o$w$CDM, $w_0w_a$CDM and o$w_0w_a$CDM.
The units of $H_0$ are $\Hunit$.
} \label{table:DEmodels}
\end{table*}

\begin{table*}
\begin{center}
\begin{tabular}{lrrrrrr}
\hline
	&	$\Omega_m$	&			$H_0$	&			$\sigma_8$	&			$\Omega_k$	&			$w$ or $w_0$	&			$w_a$			\\	\hline
$\Lambda$CDM	& $	0.308	\pm	0.010	$&$	67.8	\pm	0.7	$&$	0.815	\pm	0.009	$&$	0			$&$	-1			$&$	0			$\\	
o$\Lambda$CDM	& $	0.309	\pm	0.011	$&$	67.6	\pm	1.1	$&$	0.813	\pm	0.011	$&$	-0.001	\pm	0.003	$&$	-1			$&$	0			$\\	
$w$CDM	& $	0.311	\pm	0.022	$&$	67.6	\pm	2.4	$&$	0.813	\pm	0.023	$&$	0			$&$	-0.99	\pm	0.09	$&$	0			$\\	
o$w$CDM	& $	0.308	\pm	0.023	$&$	67.8	\pm	2.5	$&$	0.815	\pm	0.024	$&$	-0.001	\pm	0.004	$&$	-1.01	\pm	0.11	$&$	0			$\\	
$w_0w_a$CDM	& $	0.329	\pm	0.033	$&$	65.9	\pm	3.4	$&$	0.798	\pm	0.030	$&$	0			$&$	-0.78	\pm	0.29	$&$	-0.60	\pm	0.74	$\\	
o$w_0w_a$CDM	& $	0.330	\pm	0.033	$&$	65.6	\pm	3.4	$&$	0.799	\pm	0.029	$&$	-0.003	\pm	0.004	$&$	-0.75	\pm	0.28	$&$	-0.81	\pm	0.77	$\\	
\hline
\end{tabular}
\end{center}
\caption{ 
The cosmological constraints from our CMASS measurements without including observation systematics weight corrections combining with Planck data assuming $\Lambda$CDM, nonflat $\Lambda$CDM (o$\Lambda$CDM), $w$CDM, o$w$CDM, $w_0w_a$CDM and o$w_0w_a$CDM.
The units of $H_0$ are $\Hunit$.
} \label{table:DEmodels_nosys}
\end{table*}

\section{COMPARISON WITH OTHER WORKS}
\label{sec:compare}
The constraints on $H(z)r_s/r_{s,fid}$ and $D_A(z) r_{s,fid}/r_s$ are dominated by the 2-dimensional BAO feature. As shown in Fig. 13 of \cite{Anderson:2013oza}, the measurements were similar between the results from the analyses with (green) and without (blue and red) the full shape information. Note, in the same plot, the constraints from \cite{Reid:2012sw} (purple) and \cite{Sanchez:2013uxa} (black) were tighter because they either included much smaller scales or used stronger dark energy model assumption.
The recent BAO-only measurements are replying on the BAO reconstruction methodologies, e.g., see \cite{Anderson:2013zyy, Cuesta:2015mqa, Gil-Marin:2015nqa}. In those analyses, the BAO feature was enhanced but the full shape information was removed. Therefore, the information obtained from the BAO-only measurements is different from ours. BAO-only analyses do not provide $f(z)\sigma_8(z)$ measurements which could be useful for testing gravity theory, e.g., see \cite{Samushia:2012iq, Samushia:2013yga, Beutler:2013yhm, Alam:2015qta}. \cite{Gil-Marin:2015sqa} extracted the cosmological information from the full shape information using similar data sample as ours, but they performed the analysis in the Fourier space. The systematics considered in our studies have only impact on the small k-mode that they do not use. However, the nonlinear evolution and nonlinear redshift space distortion at small scales in configuration space would propagate to larger range of k-mode in Fourier space.

The redshift range used in our analysis for DR12 CMASS ($0.43 < z < 0.75$) is slightly different from the range used by \cite{Gil-Marin:2015sqa} ($0.43 < z < 0.7$). We intend to use larger volume of the sample to increase the statistics power since we drop smaller scales ($s<55$Mpc) to minimize scale-dependent effects and measure unbiased growth rate as mentioned. If we rescale our measurements to the same effective redshift z=0.57 of \cite{Gil-Marin:2015sqa}, we obtain $H(0.57)r_s = (14.29 \pm 0.48)\times10^3$ km/s, $DA(0.57)/r_s = 9.44\pm0.17$ and $f(z_{eff})\sigma_8(z_{eff})=0.488 \pm 0.060$ ($f\sigma_8$ is insensitive to the effective redshift). Despite of the different redshift range and methodology used, our measurements are in good agreement with the results from \cite{Gil-Marin:2015sqa}, $H(0.57)r_s = (13.92 \pm 0.44)\times10^3$ km/s, $D_A(0.57)/r_s = 9.42\pm0.15$, and $f(z_{eff})\sigma_8(z_{eff})=0.444 \pm 0.038$.

\begin{figure*}
\centering
\includegraphics[width=1.7 \columnwidth,clip,angle=-0]{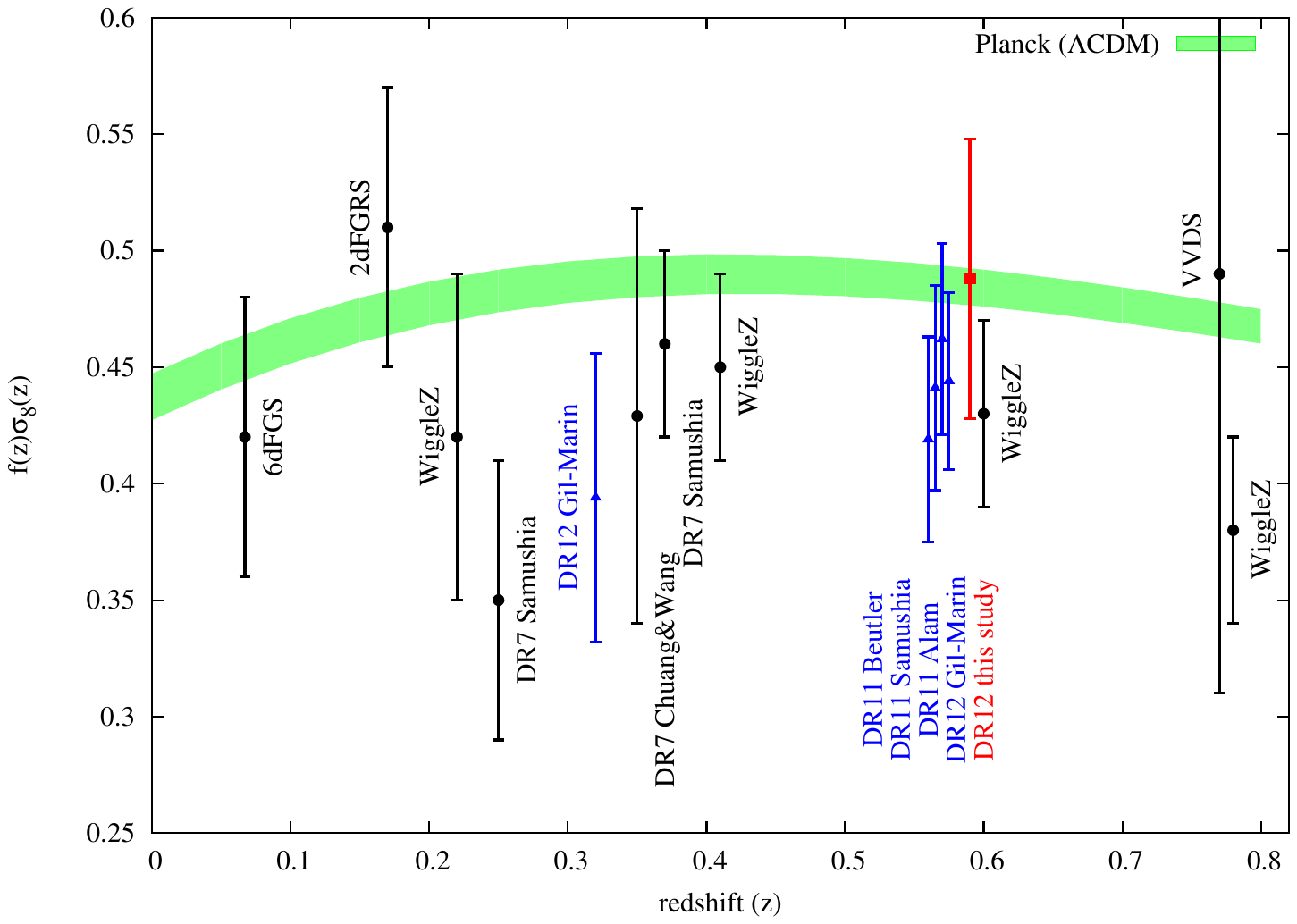}
\caption{
We compare the constraints of $f(z)\sigma_8(z)$ from CMB data (Planck) with our measurement (red square), other measurements from BOSS galaxy sample (blue triangle; \citealt{Beutler:2013yhm,Samushia:2013yga,Alam:2015qta,Gil-Marin:2015nqa}) and the measurements compiled by \protect\cite{Samushia:2012iq} (black circles). 
The constraints from CMB are obtained given $\Lambda$CDM model.
}
\label{fig:compare_fs8_lcdm}
\end{figure*}

\begin{figure*}
\centering
\includegraphics[width=1.7 \columnwidth,clip,angle=-0]{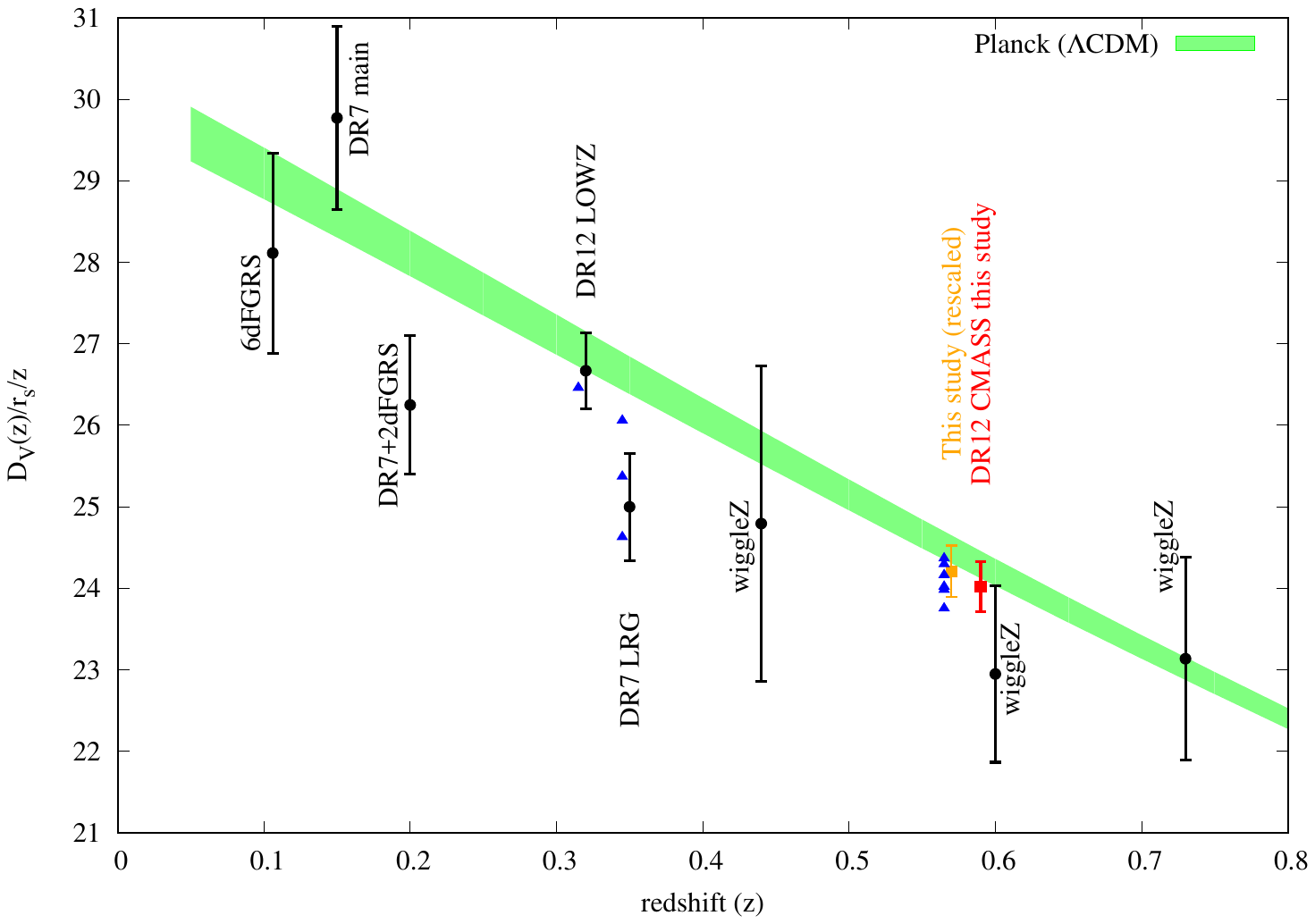}
\caption{
We compare the constraints of $\frac{D_V(z)}{r_s z}$ from CMB data (Planck) with our measurement (red square), and other measurements (black circles and blue triangles; \citealt{Blake11a,Beutler12,Percival:2009xn,Chuang:2010dv,Chuang:2011fy,Padmanabhan:2012hf,Anderson:2012sa,Anderson:2013zyy,Beutler:2013yhm,Samushia:2013yga,Tojeiro:2014eea,Ross:2014qpa}). 
When there are more than one measurements at the same redshift, we mark one of the measurement using a black circle with error bar (i.e., the measurement from \citealt{Chuang:2011fy} at $z=0.35$ and the measurement from \citealt{Cuesta:2015mqa} at $z=0.32$) and mark the others with blue triangles with slight shift in redshift to make the plot more clear. 
The constraints from CMB are obtained given $\Lambda$CDM model.
}
\label{fig:compare_dvrs_lcdm}
\end{figure*}

\begin{figure*}
\centering
\includegraphics[width=1.7 \columnwidth,clip,angle=-0]{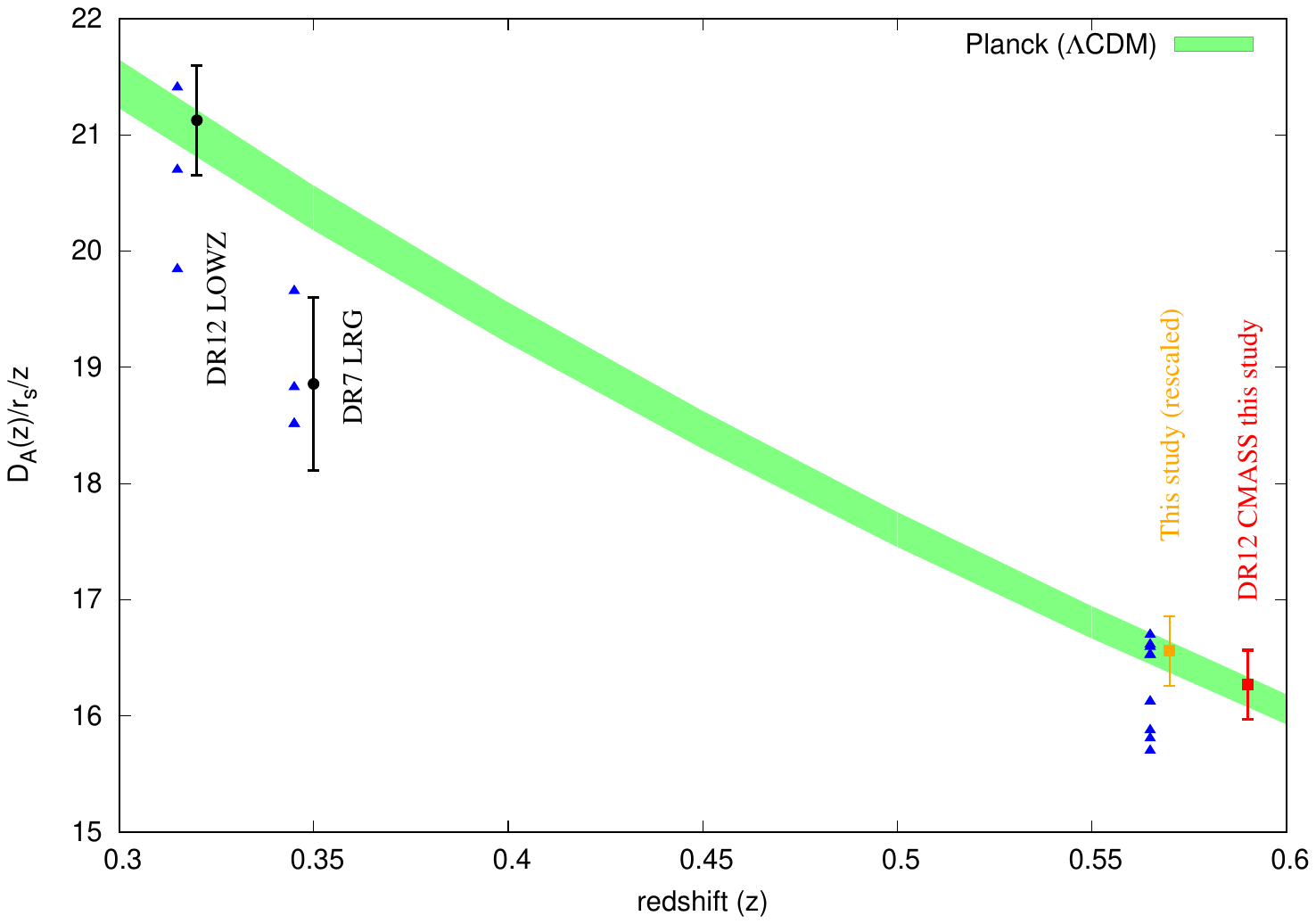}
\caption{
We compare the constraints of $\frac{D_A(z)}{r_s z}$ from CMB data (Planck) with our measurement (red square), and other measurements (black circles and blue triangles; \citealt{Chuang:2011fy,Chuang:2012ad,Chuang:2012qt,Xu:2012fw,Hemantha:2013sea,Anderson:2013oza,Chuang:2013hya, Kazin:2013rxa,Wang:2014qoa, Anderson:2013zyy,Beutler:2013yhm,Gil-Marin:2015nqa,Gil-Marin:2015sqa,Cuesta:2015mqa}). 
When there are more than one measurements at the same redshift, we mark one of the measurement using a black circle with error bar (i.e., the measurement from \citealt{Chuang:2011fy} at $z=0.35$ and the consensus value from \citealt{Gil-Marin:2015nqa,Cuesta:2015mqa} at $z=0.32$) and mark the others with blue triangles with slight shift in redshift to make the plot more clear.
The constraints from CMB are obtained given $\Lambda$CDM model.
}
\label{fig:compare_dars_lcdm}
\end{figure*}

\begin{figure*}
\centering
\includegraphics[width=1.7 \columnwidth,clip,angle=-0]{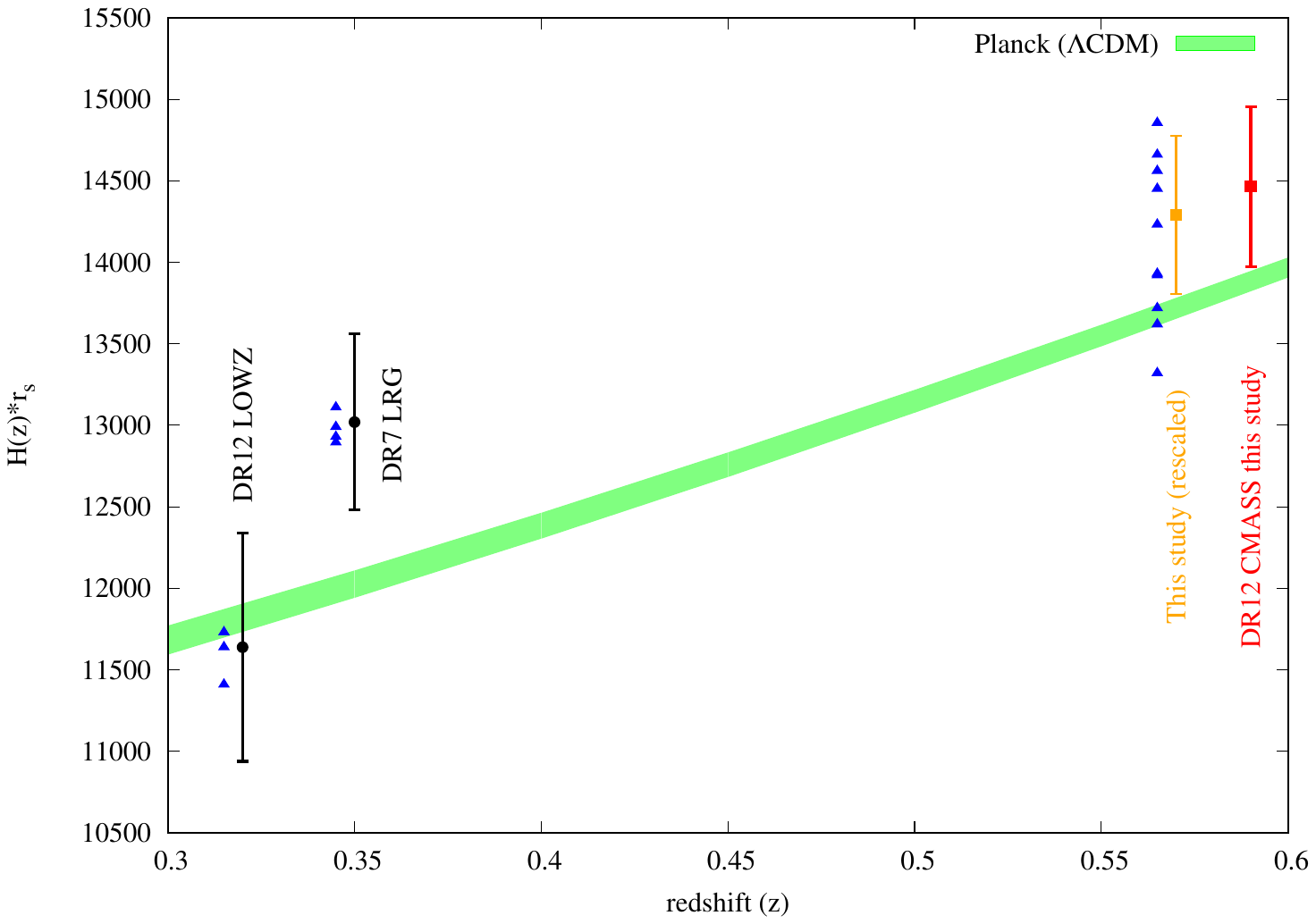}
\caption{
We compare the constraints of $H(z)r_s$ from CMB data (Planck) with our measurement (red square), and other measurements (black circles and blue triangles; \citealt{Chuang:2011fy,Chuang:2012ad,Chuang:2012qt,Xu:2012fw,Hemantha:2013sea,Anderson:2013oza,Chuang:2013hya, Kazin:2013rxa,Wang:2014qoa, Anderson:2013zyy,Beutler:2013yhm,Gil-Marin:2015nqa,Gil-Marin:2015sqa,Cuesta:2015mqa}). 
When there are more than one measurements at the same redshift, we mark one of the measurement using a black circle with error bar (i.e., the measurement from \citealt{Chuang:2011fy} at $z=0.35$ and the consensus value from \citealt{Gil-Marin:2015nqa,Cuesta:2015mqa} at $z=0.32$) and mark the others with blue triangles with slight shift in redshift to make the plot more clear.
The constraints from CMB are obtained given $\Lambda$CDM model.
}
\label{fig:compare_hrs_lcdm}
\end{figure*}

In Fig. \ref{fig:compare_fs8_lcdm}, \ref{fig:compare_dvrs_lcdm},\ref{fig:compare_dars_lcdm}, and \ref{fig:compare_hrs_lcdm},
we compare the constraints of $f(z)\sigma_8(z)$, $D_A(z)/r_s$, $H(z)r_s$, and $D_V(z)/r_s$ from CMB data (Planck assuming LCDM) with the measurements from galaxy clustering analyses.
We have included the measurements from 
VIMOS-VLT Deep Survey (VVDS;\citealt{Guzzo08}),
2dFGRS \citep{Percival:2004fs},
Six-degree-Field Galaxy Survey (6dFGS; \citealt{Beutler12}),
WiggleZ \citep{Blake11a,Blake11b},
SDSS-II/DR7 \citep{Percival:2009xn,Chuang:2010dv,Samushia:2011cs,Chuang:2011fy,Chuang:2012ad,Chuang:2012qt,Ross:2014qpa,Padmanabhan:2012hf,Xu:2012fw, 
Seo12,Hemantha:2013sea}
SDSS-III/BOSS \citep{Anderson:2012sa,Reid:2012sw, Anderson:2013oza, Chuang:2013hya, Sanchez:2013uxa, Kazin:2013rxa,Wang:2014qoa, Anderson:2013zyy,Beutler:2013yhm,Samushia:2013yga,Tojeiro:2014eea,Reid:2014iaa,Alam:2015qta,Gil-Marin:2015nqa,Gil-Marin:2015sqa,Cuesta:2015mqa}

In Fig. \ref{fig:compare_dvrs_lcdm}, \ref{fig:compare_dars_lcdm}, and \ref{fig:compare_hrs_lcdm}, when there are multiple measurements that are corresponding the same redshifts, we show the mean and error bar for one of them (indicated in the captions) and show only the means with triangles for the rest of the measurements. We also slightly shift the redshift to make the figures more clear.
One can see that the measurements of $D_V(z) r_{s,fid}/r_s$ and $f(z)\sigma_8(z)$ from different analyses but at the same redshift agree with each other. However, the measurements of $H(z)r_s/r_{s,fid}$ and $D_A(z) r_{s,fid}/r_s$ have larger scatter. This is expected since $D_V(z) r_{s,fid}/r_s$ measurement is driven by the BAO feature in the monopole and $f(z)\sigma_8(z)$ is mainly determined by the amplitude of quadrupole. But, $H(z)r_s/r_{s,fid}$ and $D_A(z) r_{s,fid}/r_s$ is correlated with the shape of BAO feature which has larger uncertainties among different models. In addition, we rescale our measurements of $H(z)r_s/r_{s,fid}$, $D_A(z) r_{s,fid}/r_s$, and $D_V(z) r_{s,fid}/r_s$, from the effective redshift $z=0.59$ (the points with red solid error bars) to $z=0.57$ (the orange points with thiner error bars) for the convenience of comparison with previous works. One can see our measurements are in agreement with others.

\section{Summary} 
\label{sec:conclusion}
We present measurements of the anisotropic galaxy clustering from the DR12 CMASS samples of the SDSS-III Baryon Oscillation Spectroscopic Survey (BOSS). 
We analyze the broad-range shape of quasi-linear scales, 
which can be modeled by perturbation theory, of the monopole and quadrupole correlation functions to obtain cosmological constraints, 
at the effective redshift $z=0.59$ of the sample, on the Hubble expansion rate $H(z)$, 
the angular-diameter distance $D_A(z)$, the normalized growth rate $f(z)\sigma_8(z)$, and the physical matter density $\Omega_mh^2$.
We obtain more robust measurements by including a polynomial as the model for the systematic errors. 
We find it works very well against the systematics effects, e.g. effects from stars and seeing. 
The parameters which are not well constrained by our
galaxy clustering analysis are marginalized over with wide flat priors.
Since no priors from other data sets (i.e., CMB) are adopted and no dark energy models are assumed, our results from BOSS CMASS galaxy clustering may be combined
with other data sets, i.e., CMB, SNe, lensing or other galaxy clustering data to constrain the parameters of a given cosmological model.
Our main results can be summarized as follows.

(i) Our measurements for DR12 CMASS ($0.43<z<0.75$), using the range $55h^{-1}$Mpc $<s<200h^{-1}$Mpc, are
$\{D_A(0.59)r_{s,fid}/r_s$, $H(0.59)r_s/r_{s,fid}$, $f(0.59)\sigma_8(0.59)$, $\Omega_m h^2\}$ = $\{1427\pm26$, $97.3\pm3.3$, $0.488 \pm 0.060$, $0.135\pm0.016\}$, where $r_s$ is the comoving sound horizon at the drag epoch and $r_{s,fid}$ is the $r_s$ of the fiducial cosmology used in this study.

(ii) In the case of the cosmological model assuming $\Lambda$CDM,
our single-probe constraints from CMASS quasi-linear scales, combined with CMB (Planck), yield the values for $\Omega_m=0.308\pm0.010$ and $H_0=67.8\pm0.7$ km$s^{-1}$Mpc$^{-1}$;
considering o$\Lambda$CDM (non-flat $\Lambda$CDM), we obtain the curvature density fraction, $\Omega_k=-0.001\pm0.003$;
adopting a constant dark energy equation of state and a flat universe ($w$CDM),
the constraint on dark energy equation of state parameter is $w=-0.98\pm0.08$.

(iii) Using our methodology and the correlation function measured without including the systematics weights corrections, we obtain the same results as the ones including the systematics weights corrections. We conclude that our measurements are robust against the known observational systematics.

\section{Acknowledgement}
C.C. would like to thank Savvas Nesseris for useful discussions.

C.C. and F.P. acknowledge support from the Spanish MICINN’s Consolider-Ingenio 2010 Programme under grant MultiDark CSD2009-00064 and AYA2010-21231-C02-01 grant.
C.C. were also supported by the Comunidad de Madrid under grant HEPHACOS S2009/ESP-1473.
MPI acknowledges support from MINECO under the grant AYA2012-39702-C02-01.

We acknowledge the use of 
the CURIE supercomputer at Tr\`es Grand Centre de calcul du CEA in France  through the French participation into the PRACE research infrastructure,
the SuperMUC supercomputer at Leibniz Supercomputing Centre of the Bavarian Academy of Science in Germany,
the TEIDE-HPC (High Performance Computing) supercomputer in Spain,
and the Hydra cluster at Instituto de F\'{\i}sica Te\'orica, (UAM/CSIC) in Spain.

Funding for SDSS-III has been provided by the Alfred P. Sloan Foundation, the Participating Institutions, the National Science Foundation, 
and the U.S. Department of Energy Office of Science. The SDSS-III web site is http://www.sdss3.org/.

SDSS-III is managed by the Astrophysical Research Consortium for the Participating Institutions of the SDSS-III Collaboration including 
the University of Arizona, the Brazilian Participation Group, Brookhaven National Laboratory, Carnegie Mellon University, University of Florida, 
the French Participation Group, the German Participation Group, Harvard University, the Instituto de Astrofisica de Canarias, 
the Michigan State/Notre Dame/JINA Participation Group, Johns Hopkins University, Lawrence Berkeley National Laboratory, Max Planck Institute for Astrophysics, 
Max Planck Institute for Extraterrestrial Physics, New Mexico State University, New York University, Ohio State University, 
Pennsylvania State University, University of Portsmouth, Princeton University, the Spanish Participation Group, University of Tokyo, University of Utah, 
Vanderbilt University, University of Virginia, University of Washington, and Yale University.

\label{lastpage}

\end{document}